\documentstyle[11pt,epsfig]{article}
\begin{document}
\title{\Large\bf  $X(1835)$ and $\eta(1760)$ observed by the BES Collaboration}

\author{\small De-Min Li \footnote{E-mail: lidm@zzu.edu.cn},~~Bing Ma\\
\small   Department of Physics, Zhengzhou University, Zhengzhou,
Henan 450052, P. R. China}
\date{\today}
\maketitle
\vspace{0.5cm}

\begin{abstract}

With the assumption that the $X(1835)$ and $\eta(1760)$ recently
observed by the BES Collaboration are the $3\,^1S_0$ meson states,
the strong decays of these two states are investigated in the
$^3P_0$ decay model. We find that the predicted total widths of
the $X(1835)$ and $\eta(1760)$ can be reasonably reproduced with
the $X(18350)-\eta(1760)$ mixing angle lying on the range from
$-0.26$ to $+0.55$ radians. Further, the mixing angle of the
$X(1835)$ and $\eta(1760)$ is phenomenologically determined to be
about $-0.24$ radians in the presence of the $\pi(1800)$,
$K(1830)$, $\eta(1760)$ and $X(1835)$ belonging to the $3\,^1S_0$
meson nonet. Our estimated mixing angle can naturally account for
the widths of the $X(1835)$ and $\eta(1760)$, which shows that the
assignment of the $X(1835)$ and $\eta(1760)$, together the
$\pi(1800)$ and $K(1830)$ as the $3\,^1S_0$ $q\bar{q}$ members
seems reasonable.

\end{abstract}

\vspace{0.5cm}


\newpage

\baselineskip 24pt

\section{Introduction}
\indent \vspace*{-1cm}

In 2005, the BES collaboration observed a narrow peak called
$X(1835)$ in the reaction
$J/\psi\rightarrow\gamma\pi^+\pi^-\eta^\prime$ with a statistical
significance of $7.7\sigma$, and the mass and width of the
$X(1835)$ obtained from the fit with a Breit-Wigner function are
$M=(1833.7\pm 6.1\pm 2.7)$ MeV and $\Gamma=(67.7\pm 20.3\pm 7.7)$
MeV, respectively\cite{x1835}. No partial wave analysis has been
made but based on production and decay, it is most likely that the
$J^{PC}I^G$ of the $X(1835)$ is $0^{-+}0^+$.

The BES collaboration suggests that the $X(1835)$ is related to
the $p\bar{p}$ mass threshold enhancement observed in the
$J/\psi\rightarrow\gamma p\bar{p}$ channel\cite{pp}, which
therefore triggering many exotic speculations of the nature of the
$X(1835)$ such as glueball\cite{g1,g2,g3}, $p\bar{p}$
baryonium\cite{b1,b2} etc. It should be noted that there is no
strong experimental evidence that the $p\bar{p}$ threshold
enhancement and the $X(1835)$ are the same resonance, as pointed
out by Huang and Zhu\cite{hz}; very probably they have completely
different underlying structures. In the presence of the $X(1835)$
being not related to the $p\bar{p}$ mass threshold enhancement,
Huang and Zhu suggested the $X(1835)$ is the second radial
excitation of the $\eta^\prime$\cite{hz}, while Klempt and Zaitsev
rather believed the $X(1835)$ to be the $\eta^\prime$'s first
radial excitation\cite{KZ}.

The $\eta(1760)$ was reported by the MARKIII Collaboration in
radiative $J/\psi$ decays into $\omega\omega$\cite{MKo1} and
$\rho\rho$\cite{MKr1}. It was also observed by the DM2
Collaboration in $J/\psi$ radiative decays in the
$\rho\rho$\cite{DM2r1} decay mode with a mass of $M=(1760\pm 11)$
MeV and a width of $\Gamma=(60\pm 16)$ MeV and in the
$\omega\omega$ decay mode\cite{DM2o1}. Vijande et al. suggested
that it can be identified as the $2\,^1S_0$ $q\bar{q}$
state\cite{vij}; Page and Li proposed that it could have hybrid
admixture\cite{pageli}. A reanalysis MARKIII data on
$J/\psi\rightarrow \gamma 4\pi$ performed by Bugg et al. suggested
that the decay mode should not be in the $\rho\rho$ pseudoscalar
wave but in the
$(\pi\pi)_{\mbox{\tiny{S-wave}}}(\pi\pi)_{\mbox{\tiny {S-wave}}}$
$0^{++}$ scalar wave\cite{Bugg}. This was supported by BES
Collaboration suggesting the $\rho\rho$ resonance at $1760$ MeV
should be interpreted as the scalar meson in its $\sigma\sigma$
decay\cite{BES17601}. The conclusions regarding the presence of
this pseudoscalar signal in the MARKIII $4\pi$ data become
therefore controversial.

More Recently, the decay channel $J/\psi\rightarrow \gamma
\omega\omega,~\omega\rightarrow\pi^+\pi^-\pi^0$ was analyzed by
the BES collaboration, and a strong enhancement was found in the
$\omega\omega$ invariant mass distribution at $1.76$ GeV. The
partial wave analysis indicated that the structure was
predominantly pseudoscalar with small scalar and tensor
contributions. The mass of the pseudoscalar state also called
$\eta(1760)$ is $M=(1744\pm 10\pm 15)$ MeV, the width
$\Gamma=(244^{+24}_{-21}\pm 25) $ MeV and the product branching
fraction is $\mbox{Br}(J/\psi\rightarrow
\gamma\eta(1760))\mbox{Br}(\eta(1760)\rightarrow
\omega\omega)=(1.98\pm 0.08\pm 0.32)\times
10^{-3}$\cite{BES17602}. The BES Collaboration suggested that it
could be a mixture of the glueball and the $q\bar{q}$ meson.

From PDG2006\cite{pdg2006}, the $1\,^1S_0$ meson nonet ($\pi$,
$\eta$, $\eta^\prime$ and $K$) as well as the $2\,^1S_0$ members
[$\pi(1300)$, $\eta(1295)$ and $\eta(1475)$] have been well
established. We argue that the $X(1835)$ and $\eta(1760)$ reported
by BES Collaboration can be assigned as the ordinary $3\,^1S_0$
$q\bar{q}$ states. The purpose of this work is to check in the
simple picture of the $X(1835)$ and $\eta(1760)$ as the $3\,^1S_0$
$q\bar{q}$ states, whether the total widths of these two states
can be reasonably reproduced in the $^3P_0$ meson decay model or
not.

The organization of this paper is as follows. In section 2, the
brief review of the $^3P_0$ decay model is given (For the detailed
review see {\sl e.g.}
Refs.\cite{3p0rev1,3p0rev2,3p0rev3,3p0rev4}.) In section 3, the
decay widths of the $X(1835)$ and $\eta(1760)$ are presented. The
mixing angle of the $X(1835)-\eta(1760)$ is given in section 4,
and a summary and conclusion are given in section 5.

\section{ The $^3P_0$ meson decay model}
\indent \vspace*{-1cm}

 The $^3P_0$ decay model, also known as the quark-pair creation
model, was originally introduced by Micu\cite{micu} and further
developed by Le Yaouanc et al.\cite{3p0rev1}. The $^3P_0$ decay
model has been widely used to evaluate the strong decays of
hadrons\cite{3p00,3p0y,3p0x,3p0x1,3p0x2,3p01,3p02,3p03,quarkmass,3p04},
since it gives a good description of many of the observed decay
amplitudes and partial widths of the hadrons. The main assumption
of the $^3P_0$ decay model is that strong decays take place via
the creation of a $^3P_0$ quark-antiquark pair from the vacuum.
The new produced quark-antiquark pair, together with the
$q\bar{q}$ within the initial meson regroups into two outgoing
mesons in all possible quark rearrangement ways, which corresponds
to the two decay diagrams as shown in Fig.1 for the meson decay
process $A\rightarrow B+C$.

\begin{figure}[hbt]
\begin{center}
\epsfig{file=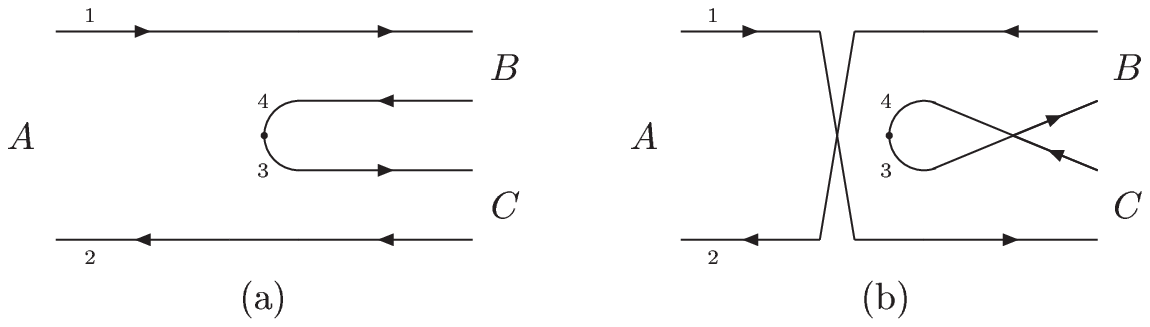,width=12.0cm, clip=}
\vspace*{0.5cm}\vspace*{-1cm}
 \caption{\small The two possible diagrams contributing to $A\rightarrow B+C$ in the $^3P_0$
 model.}
\end{center}
\end{figure}

The transition operator $T$ of the decay $A\rightarrow BC$ in the
$^3P_0$ model is given by
\begin{eqnarray}
T=-3\gamma\sum_m\langle 1m1-m|00\rangle\int
d^3\vec{p}_3d^3\vec{p}_4\delta^3(\vec{p}_3+\vec{p}_4){\cal{Y}}^m_1(\frac{\vec{p}_3-\vec{p}_4}{2})\chi^{34}_{1-m}\phi^{34}_0\omega^{34}_0b^\dagger_3(\vec{p}_3)d^\dagger_4(\vec{p}_4),
\end{eqnarray}
where $\gamma$ is a dimensionless parameter representing the
probability of the quark-antiquark pair $q_3\bar{q}_4$ with
$J^{PC}=0^{++}$ creation from the vacuum, $\vec{p}_3$ and
$\vec{p}_4$ are the momenta of the created quark $q_3$ and
antiquark  $\bar{q}_4$, respectively. $\phi^{34}_{0}$,
$\omega^{34}_0$, and $\chi_{{1,-m}}^{34}$ are the flavor, color,
and spin wave functions of the  $q_3\bar{q}_4$, respectively. The
solid harmonic polynomial
${\cal{Y}}^m_1(\vec{p})\equiv|p|^1Y^m_1(\theta_p,\phi_p)$ reflects
the momentum-space distribution of the $q_3\bar{q}_4$ .

For the meson wave function, we adopt the mock meson
$|A(n_A{}^{2S_A+1}L_{A}\,\mbox{}_{J_A M_{J_A}})(\vec{P}_A)\rangle$
defined by\cite{mock}
\begin{eqnarray}
|A(n_A{}^{2S_A+1}L_{A}\,\mbox{}_{J_A M_{J_A}})(\vec{P}_A)\rangle
&\equiv& \sqrt{2E_A}\sum_{M_{L_A},M_{S_A}}\langle L_A M_{L_A} S_A
M_{S_A}|J_A
M_{J_A}\rangle\nonumber\\
&&\times  \int d^3\vec{p}_A\psi_{n_AL_AM_{L_A}}(\vec{p}_A)\chi^{12}_{S_AM_{S_A}}\phi^{12}_A\omega^{12}_A\nonumber\\
&&\times  |q_1({\scriptstyle
\frac{m_1}{m_1+m_2}}\vec{P}_A+\vec{p}_A)\bar{q}_2
({\scriptstyle\frac{m_2}{m_1+m_2}}\vec{P}_A-\vec{p}_A)\rangle,
\end{eqnarray}
where $m_1$ and $m_2$ are the masses of the quark $q_1$ with a
momentum of $\vec{p}_1$ and the antiquark $\bar{q}_2$ with a
momentum of $\vec{p}_2$, respectively. $n_A$ is the radial quantum
number of the meson $A$ composed of $q_1\bar{q}_2$.
$\vec{S}_A=\vec{s}_{q_1}+\vec{s}_{q_2}$,
$\vec{J}_A=\vec{L}_A+\vec{S}_A$, $\vec{s}_{q_1}$ ($\vec{s}_{q_2}$)
is the spin of $q_1$ ($q_2$), $\vec{L}_A$ is the relative orbital
angular momentum between $q_1$ and $q_2$.
$\vec{P}_A=\vec{p}_1+\vec{p}_2$,
$\vec{p}_A=\frac{m_1\vec{p}_1-m_1\vec{p}_2}{m_1+m_2}$. $\langle
L_A M_{L_A} S_A M_{S_A}|J_A M_{J_A}\rangle$ is a Clebsch-Gordan
coefficient, and  $E_A$ is the total energy of the meson $A$.
$\chi^{12}_{S_AM_{S_A}}$, $\phi^{12}_A$, $\omega^{12}_A$, and
$\psi_{n_AL_AM_{L_A}}(\vec{p}_A)$ are the spin, flavor, color, and
space wave functions of the meson $A$, respectively. The mock
meson satisfies the normalization condition
\begin{eqnarray}
\langle A(n_A{}^{2S_A+1}L_{A}\,\mbox{}_{J_A M_{J_A}})(\vec{P}_A)
|A(n_A{}^{2S_A+1}L_{A}\,\mbox{}_{J_A
M_{J_A}})(\vec{P^\prime}_A)\rangle=2E_A\delta^3(\vec{P}_A-\vec{P^\prime}_A).
\end{eqnarray}
The $S$-matrix of the process $A\rightarrow BC$ is defined by
\begin{eqnarray}
\langle BC|S|A\rangle=I-2\pi i\delta(E_A-E_B-E_C)\langle
BC|T|A\rangle,
\end{eqnarray}
with
\begin{eqnarray}
\langle
BC|T|A\rangle=\delta^3(\vec{P}_A-\vec{P}_B-\vec{P}_C){\cal{M}}^{M_{J_A}M_{J_B}M_{J_C}},
\end{eqnarray}
where ${\cal{M}}^{M_{J_A}M_{J_B}M_{J_C}}$ is the helicity
amplitude of $A\rightarrow BC$. In the center of mass frame of
meson $A$, ${\cal{M}}^{M_{J_A}M_{J_B}M_{J_C}}$ can be written as
\begin{eqnarray}
{\cal{M}}^{M_{J_A}M_{J_B}M_{J_C}}(\vec{P})&=&\gamma\sqrt{8E_AE_BE_C}
\sum_{\renewcommand{\arraystretch}{.5}\begin{array}[t]{l}
\scriptstyle M_{L_A},M_{S_A},\\\scriptstyle M_{L_B},M_{S_B},\\
\scriptstyle M_{L_C},M_{S_C},m
\end{array}}\renewcommand{\arraystretch}{1}\!\!
\langle L_AM_{L_A}S_AM_{S_A}|J_AM_{J_A}\rangle\nonumber\\
&&\times\langle L_BM_{L_B}S_BM_{S_B}|J_BM_{J_B}\rangle\langle
L_CM_{L_C}S_CM_{S_C}|J_CM_{J_C}\rangle\nonumber\\
&&\times\langle 1m1-m|00\rangle\langle
\chi^{14}_{S_BM_{S_B}}\chi^{32}_{S_CM_{S_C}}|\chi^{12}_{S_AM_{S_A}}\chi^{34}_{1-m}\rangle\nonumber\\
&&\times[f_1I(\vec{P},m_1,m_2,m_3)+(-1)^{1+S_A+S_B+S_C}f_2I(-\vec{P},m_2,m_1,m_3)],
\end{eqnarray}
with $f_1=\langle
\phi^{14}_B\phi^{32}_C|\phi^{12}_A\phi^{34}_0\rangle$ and $f_2=
\langle\phi^{32}_B\phi^{14}_C|\phi^{12}_A\phi^{34}_0\rangle$,
corresponding to the contributions from Figs. 1 (a) and 1 (b),
respectively, and
\begin{eqnarray} I(\vec{P},m_1,m_2,m_3)&=&\int
d^3\vec{p}\,\mbox{}\psi^\ast_{n_BL_BM_{L_B}}
({\scriptstyle\frac{m_3}{m_1+m_2}}\vec{P}_B+\vec{p})\psi^\ast_{n_CL_CM_{L_C}}
({\scriptstyle\frac{m_3}{m_2+m_3}}\vec{P}_B+\vec{p})\nonumber\\
&&\times\psi_{n_AL_AM_{L_A}}
(\vec{P}_B+\vec{p}){\cal{Y}}^m_1(\vec{p}),
\end{eqnarray}
where $\vec{P}=\vec{P}_B=-\vec{P}_C$, $\vec{p}=\vec{p}_3$, $m_3$
is the mass of the created quark $q_3$.

The spin overlap in terms of Winger's $9j$ symbol can be given by
\begin{eqnarray}
&&\langle
\chi^{14}_{S_BM_{S_B}}\chi^{32}_{S_CM_{S_C}}|\chi^{12}_{S_AM_{S_A}}\chi^{34}_{1-m}\rangle=\nonumber\\
&&\sum_{S,M_S}\langle S_BM_{S_B}S_CM_{S_C}|SM_S\rangle\langle
S_AM_{S_A}1-m|SM_S\rangle\nonumber\\
&&(-1)^{S_C+1}\sqrt{3(2S_A+1)(2S_B+1)(2S_C+1)}\left\{\begin{array}{ccc}
\frac{1}{2}&\frac{1}{2}&S_A\\
\frac{1}{2}&\frac{1}{2}&1\\
S_B&S_C&S
\end{array}\right\}.
\end{eqnarray}

 In order to compare with experiment conventionally,
${\cal{M}}^{M_{J_A}M_{J_B}M_{J_C}}(\vec{P})$ can be converted into
the partial amplitude by a recoupling calculation\cite{recp}
\begin{eqnarray}
{\cal{M}}^{LS}(\vec{P})&=&
\sum_{\renewcommand{\arraystretch}{.5}\begin{array}[t]{l}
\scriptstyle M_{J_B},M_{J_C},\\\scriptstyle M_S,M_L
\end{array}}\renewcommand{\arraystretch}{1}\!\!
\langle LM_LSM_S|J_AM_{J_A}\rangle\langle
J_BM_{J_B}J_CM_{J_C}|SM_S\rangle\nonumber\\
&&\times\int
d\Omega\,\mbox{}Y^\ast_{LM_L}{\cal{M}}^{M_{J_A}M_{J_B}M_{J_C}}
(\vec{P}).
\end{eqnarray}
If we consider the relativistic phase space, the decay width
$\Gamma(A\rightarrow BC)$ in terms of the partial wave amplitudes
is
\begin{eqnarray}
\Gamma(A\rightarrow BC)= \frac{\pi
P}{4M^2_A}\sum_{LS}|{\cal{M}}^{LS}|^2. \label{width1}
\end{eqnarray}
Here
$P=|\vec{P}|$=$\frac{\sqrt{[M^2_A-(M_B+M_C)^2][M^2_A-(M_B-M_C)^2]}}{2M_A}$,
$M_A$, $M_B$, and $M_C$ are the masses of the meson $A$, $B$, and
$C$, respectively.

The decay width can be derived analytically if the simple harmonic
oscillator (SHO) approximation for the meson space wave functions
is used. In momentum-space, the SHO wave function is
\begin{eqnarray}
\psi_{nLM_L}(\vec{p})=R^{\mbox{\tiny
SHO}}_{nL}(p)Y_{LM_L}(\Omega_p),
\end{eqnarray}
where the radial wave function is given by
\begin{eqnarray}
R^{\mbox{\tiny SHO}}_{nL}=\frac{(-1)^n(-i)^L}{\beta^{\frac{3}{2}}}
\sqrt{\frac{2n!}{\Gamma(n+L+\frac{3}{2})}}\left(\frac{p}{\beta}\right
)^L e^{-\frac{p^2}{2\beta^2}}L^{L+\frac{1}{2}}_n({\scriptstyle
\frac{p^2}{\beta^2}}).
\end{eqnarray}
Here $\beta$ is the SHO wave function scale parameter, and
$L^{L+\frac{1}{2}}_n({\scriptstyle \frac{p^2}{\beta^2}})$ is  an
associated Laguerre polynomial.

The SHO wave functions can not be regarded as realistic, however,
they are a {\it {de facto}} standard for many nonrelativistic
quark model calculations. Moreover, the more realistic space wave
functions such as those obtained from Coulomb plus the linear
potential model do not always result in systematic improvements
due to the inherent uncertainties of the $^3P_0$ decay model
itself\cite{3p0y,3p0x,3p0x2}. The SHO wave function approximation
is commonly employed in the $^3P_0$ decay model in literature. In
the present work, the SHO wave function approximation for the
meson space wave functions is taken.

\section{ Decays of the $X(1835)$ and $\eta(1760)$ in the $^3P_0$
model}
\indent \vspace*{-1cm}

It is well known that in a meson nonet, the physical isoscalar
states can mix. With the assumption that the $X(1835)$ and
$\eta(1760)$ being the $3\,^1S_0$ meson nonet,  the
$X(1835)-\eta(1760)$ mixing can be parameterized as
\begin{eqnarray}
&&\eta(1760)=\cos\phi n\bar{n}-\sin\phi s\bar{s},\\
&&X(1835)=\sin\phi n\bar{n}+\cos\phi s\bar{s},
\end{eqnarray}
where $n\bar{n}=(u\bar{u}+d\bar{d})/\sqrt{2}$ and $s\bar{s}$ are
the pure $3\,^1S_0$ nonstrange and strange states, respectively.

According to (\ref{width1}), the partial widths of $X(1835)$ and
$\eta(1760)$ become with mixing
\begin{eqnarray}
&&\Gamma(\eta(1760)\rightarrow
BC)=\frac{\pi~P}{4M^2_{\eta(1760)}}\sum_{LS}|\cos\phi
{\cal{M}}^{LS}_{n\bar{n}\rightarrow BC}-\sin\phi
{\cal{M}}^{LS}_{s\bar{s}\rightarrow BC}|^2,
\label{w1}\\
 &&\Gamma(X(1835)\rightarrow
BC)=\frac{\pi~P}{4M^2_{\eta(1835)}}\sum_{LS}|\sin\phi
{\cal{M}}^{LS}_{n\bar{n}\rightarrow BC}+\cos\phi
{\cal{M}}^{LS}_{s\bar{s}\rightarrow BC}|^2. \label{w2}
\end{eqnarray}

Under the SHO wave function approximation, the parameters used in
the $^3P_0$ decay model involve the $q\bar{q}$ pair production
strength parameter $\gamma$, the SHO wave function scale parameter
$\beta$, and the masses of the constituent quarks. In the present
work, we take $\gamma=6.95$ and $\beta=0.4$ GeV, the typical
values used to evaluate the light meson
decays\cite{3p0x,3p0x1,3p0x2,3p01,3p02,3p03}\footnote{Our value of
$\gamma$ is higher than that used by other groups such as
\cite{3p0x2,3p01,3p02,3p03} by a factor of $\sqrt{96\pi}$ due to
different field conventions, constant factor in $T$, etc. The
calculated results of the widths are, of course, unaffected.}, and
$m_u=m_d=0.33$ GeV, $m_s=0.55$ GeV\cite{quarkmass}. Based on the
partial wave amplitudes of the $3\,^1S_0$ $q\bar{q}$ into two
other mesons listed in the Appendix A and the flavor and charge
multiplicity factors shown in Table 2, from (\ref{w1}) and
(\ref{w2}), the numerical values of the partial decay widths of
the $\eta(1760)$ and $X(1835)$ are listed in Table 1. Masses of
the final mesons are taken from PDG2006\cite{pdg2006}. We assume
the $a_0(1450)$ is the ground scalar mesons as
Ref.\cite{3p02,3p03}. The total widths of the $X(1835)$ and
$\eta(1760)$ are shown in Fig. 2 as functions of the mixing angle
$\phi$.

\begin{table}[hbt]
\begin{center}
\caption{\small Decays of the $X(1835)$ and $\eta(1760)$ as the
$3\,^1S_0$ $q\bar{q}$ in the $^3P_0$ model. $c\equiv\cos\phi$,
$s\equiv\sin\phi$.}
\vspace*{0.5cm}
\begin{tabular}{c|c|c}\hline\hline
         &\multicolumn{1}{c}{$\eta(1760)$}& \multicolumn{1}{c}{$X(1835)$}
         \\  \hline
   Mode  & $\Gamma_i$(MeV) &    $\Gamma_i$(MeV)
   \\\hline
$\rho\rho$ &
  $92.7c^2$&$117.1s^2$\\
$\omega\omega$&
  $29.4c^2$&$38.7s^2$\\
$a_2(1320)\pi$&
  $33.1c^2$&$82.3s^2$\\
$a_0(1450)\pi$&
 $26.7c^2$&$30.0s^2$\\
$KK^\ast$&
  $51.8c^2+93.1cs+41.8s^2$&$27.8c^2-83.6cs+62.8s^2$\\
$K^\ast K^\ast$&
  &$16.6c^2+21.3cs+6.8s^2$\\\hline
&\multicolumn{1}{c}{$\Gamma=233.7c^2+93.1cs+41.8s^2$}&
\multicolumn{1}{|c}{$\Gamma=44.4c^2-62.3cs+337.7s^2$}\\
&\multicolumn{1}{c}{$\Gamma_{\mbox{expt}}=244^{+24}_{-21}\pm 25$}&
\multicolumn{1}{|c}{$\Gamma_{\mbox{expt}}=67.7\pm 20.3\pm7.7$}
         \\  \hline
 \hline
\end{tabular}
\end{center}
\end{table}

\vspace*{1cm}

 \begin{figure}[hbt]
 \begin{center}
\vspace{-1.5cm} \epsfig{file=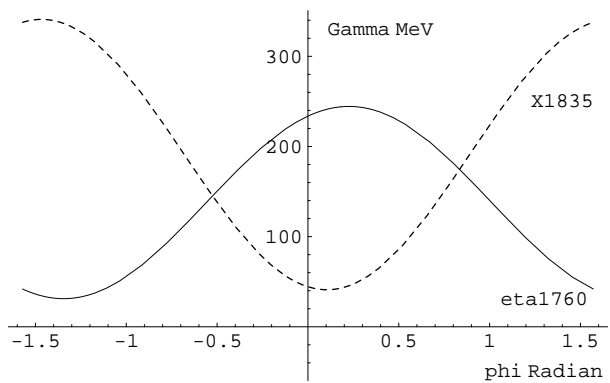,width=8.0cm, clip=}
\vspace*{0.5cm}\vspace*{-1cm}
 \caption{\small Theoretical total widths of the $X(1835)$ and $\eta(1760)$ versus the mixing angle $\phi$
 }
\end{center}
\end{figure}

From Fig. 2, we find that in the presence of the
$X(18350)-\eta(1760)$ mixing angle $\phi$ lying on the range from
about $-0.26$ to $+0.55$ radians, both $\Gamma(X(1835))$ and
$\Gamma(\eta(1760))$ can be reasonably reproduced. In order to
check whether the possibility of $-0.26 \leq \phi \leq +0.55$
radians exists or not, below we shall estimate the
$X(1835)$-$\eta(1760)$ mixing angle phenomenologically.

\section{The $X(1835)$-$\eta(1760)$ mixing angle}
\indent \vspace*{-1cm}

 In
the $n\bar{n}$ and $s\bar{s}$ basis, the mass-squared matrix
describing the $X(1835)$ and $\eta(1760)$ mixing can be written
as\cite{jpglidm,epja-1}
\begin{eqnarray}
M^2=\left(\begin{array}{cc}
M^2_{n\bar{n}}+2A_m&\sqrt{2}A_mX\\
\sqrt{2}A_mX&M_{s\bar{s}}^2+A_mX^2
\end{array}\right),
\label{matrix}
\end{eqnarray}
where $M_{n\bar{n}}$ and $M_{s\bar{s}}$ are the masses of the
states $n\bar{n}$ and $s\bar{s}$, respectively, $A_m$ denotes the
total annihilation strength of the $q\bar{q}$ pair for the light
flavors $u$ and $d$, $X$ describes the $SU(3)$-breaking ratio of
the nonstrange and strange quark propagators via the constituent
quark mass ratio $m_u/m_s$. The masses of the two physical states
$\eta(1760)$ and $X(1835)$ can be related to the matrix $M^2$ by
the unitary matrix $U=\left(\begin{array}{cc}
\cos\phi&-\sin\phi\\
\sin\phi&\cos\phi\end{array}\right)$
\begin{eqnarray}
U M^2 U^\dagger=\left(\begin{array}{cc}
M^2_{\eta(1760)}&0\\
0&M^2_{X(1835)}\end{array}\right). \label{diag}
\end{eqnarray}

$n\bar{n}$ is the orthogonal partner of $\pi(3\,^1S_0)$, the
isovector state of $3\,^1S_0$ meson nonet, and one can expect that
$n\bar{n}$ degenerates with $\pi(3\,^1S_0)$ in effective quark
masses, here we take
$M_{n\bar{n}}=M_{\pi(3\,^1S_0)}=M_{\pi(1800)}$\footnote{The nature
of the $\pi(1800)$ is controversial, different interpretations
such as a $3\,^1S_0$ $q\bar{q}$ [$\pi(3\,^1S_0)$] and hybrid
($\pi_H$) exist\cite{KZ}. There is the possibility that the two
states, $\pi(3\,^1S_0)$ and $\pi_H$ have been observed in the 1800
MeV mass region, as pointed out by\cite{3p02}. Here we consider
the $\pi(1800)$ as the $\pi(3\,^1S_0)$, as suggested by
PDG2002\cite{pdg2002}}. From the masses of the constituent quarks
used to evaluate the widths of the $\eta(1760)$ and $X(1835)$ in
section 3, we have $X=0.33/0.55=0.6$. Inputting the masses of the
$\pi(1800)$, $\eta(1760)$ and $X(1835)$, with the help of the
Gell-Mann-Okubo mass formula $M^2_{s\bar{s}}=2M^2_{
K(3\,^1S_0)}-M^2_{n\bar{n}}$\cite{okubo}, from relation
(\ref{diag}) we can have $A_m=-0.09~ {\mbox{GeV}}^2$ and
$M_{K(3\,^1S_0)}=1.82 $ GeV in agreement with the experimental
result $M_{K(1830)}\sim 1830$ MeV\cite{pdg2006}. The predicted
total width $\Gamma_{K(3\,^1S_0)}$ given by the $^3P_0$ decay
model is about 201 MeV\cite{3p03}, also roughly compatible with
the experimental result $\Gamma(1830)\sim 250$ MeV\cite{pdg2006}.
Therefore, in the presence of the $\pi(1800)$, $\eta(1760)$ and
$X(1835)$ being the $3\,^1S_0$ states, the $K(1830)$ seems an
excellent candidate for the second radial excitation of the
Kaon\cite{pdg2002}.

Based on the values of the above parameters involved in
(\ref{matrix}), we have
\begin{eqnarray}
U=\left(\begin{array}{cc}
\cos\phi&-\sin\phi\\
\sin\phi&\cos\phi\end{array}\right)=\left(\begin{array}{cc}
 +0.97&+0.24\\
-0.24&+0.97
\end{array}\right),
\label{mixangle}
\end{eqnarray}
which gives $\phi=-0.24$ radians, just lying on the range of about
$-0.26\sim +0.55$ radians.  From Table 1, this estimated mixing
angle leads to $\Gamma(1760)=200.6$ MeV and $\Gamma(1835)=75.7$
MeV, both in agreement with the experimental results within
errors. This shows that in the presence of the $\pi(1800)$,
$K(1830)$, $X(1835)$ and $\eta(1760)$ belonging to the $3\,^1S_0$
meson nonet, the total widths of the $X(1835)$ and $\eta(1760)$
can be naturally accounted for in the $^3P_0$ decay model.
Therefore, the assignment of the $X(1835)$ and $\eta(1760)$,
together the $\pi(1800)$ and $K(1830)$ as the $3\,^1S_0$
$q\bar{q}$ members seems reasonable.

As mentioned in section 1, in $J/\psi$ radiative decays, the DM2
Collaboration observed the $\rho\rho$ signal with a mass of
$(1760\pm 11)$ MeV and a width of $\Gamma=(60\pm 16)$
MeV\cite{DM2r1}, while the BES Collaboration found the
$\omega\omega$ signal with a mass of $(1744\pm 10\pm 15)$ MeV and
a width of $\Gamma=(244^{+24}_{-21}\pm 25) $ MeV\cite{BES17602}.
It is suggested that the $\rho\rho$ signal is compatible with the
$\omega\omega$ signal\cite{KZ,pdg2006}; we regard them as
incompatible. The $^3P_0$ decay model predicts that
$\frac{{\cal{B}}(\eta(1760)\rightarrow\rho\rho)}{{\cal{B}}(\eta(1760)\rightarrow\omega\omega)}=3$,
i.e, the $\rho\rho$ yield should be $3$ times larger than that for
$\omega\omega$, incompatible with the measured yields $(1.44\pm
0.12\pm 0.21)\times 10^{-3}$ for $\rho\rho$\cite{DM2r1} and
$(1.98\pm 0.08\pm 0.32)\times 10^{-3}$ for
$\omega\omega$\cite{BES17602}, which indicates that the $\rho\rho$
signal at $1760$ MeV\cite{DM2r1} may be incompatible with the
$\omega\omega$ signal at $1760$ MeV\cite{BES17602}, assuming that
the $^3P_0$ decay model is accurate.

Table 1 and (\ref{mixangle}) show that the modes $K^\ast K$ and
$K^\ast K^\ast$ are the dominant decay modes of the $X(1835)$.
These two modes are experimental attractive because their
$\cos\phi\sin\phi$ cross terms have opposite signs, so the ratio
${\cal{B}}(X(1835)\rightarrow K^\ast
K)$/${\cal{B}}(X(1835)\rightarrow K^\ast K^\ast)$ depends strongly
on the mixing angle $\phi$.

\section{Summary and conclusion}
\indent \vspace*{-1cm}

Assuming that the $X(1835)$ and $\eta(1760)$ reported by the BES
Collaboration are the ordinary $3\,^1S_0$ $q\bar{q}$ states, we
evaluate the strong decays of these two states in the framework of
the $^3P_0$ meson decay model. We find that when  the
$X(18350)-\eta(1760)$ mixing angle lies on the range from $-0.26$
to $+0.55$ radians, both the total width of the $X(1835)$ and that
of the $\eta(1760)$ can be reasonably reproduced. Also, in the
presence of the $\pi(1800)$, $\eta(1760)$ and $X(1835)$ belonging
to the $3\,^1S_0$ meson nonet, the $K(1830)$ seems an excellent
candidate for the $3\,^1S_0$ kaon, and the $X(1835)-\eta(1760)$
mixing angle of about $-0.24$ radians can be phenomenologically
obtained, naturally accounting for the total widths of the
$X(1835)$ and $\eta(1760)$. We therefore conclude that the
assignment of the $X(1835)$ and $\eta(1760)$, together the
$\pi(1800)$ and $K(1830)$ as the $3\,^1S_0$ $q\bar{q}$ members
seems reasonable, and the $X(1835)$ is mostly strange while the
$\eta(1760)$ is mainly non-strange. Also, we suggest the
$\rho\rho$ signal at $1760$ MeV reported by the DM2
Collaboration\cite{DM2r1} may be incompatible with the
$\omega\omega$ signal at $1760$ MeV observed by the BES
Collaboration.

 \section*{Acknowledgments}
 This work
is supported in part by HANCET under Contract No. 2006HANCET-02
and Program for Youthful Teachers in University of Henan Province.

\appendix
\newcounter{zaehler}
\renewcommand{\thesection}{\Alph{zaehler}}
\renewcommand{\theequation}{\Alph{zaehler}.\arabic{equation}}
\setcounter{equation}{0} \addtocounter{zaehler}{1}
\section*{Appendix A: The amplitudes for the $3\,^1S_0$ $q\bar{q}$ decay in ${^3P_0}$
model} \vspace*{-0.8cm} {\tiny
\begin{eqnarray}
&&{\cal{M}}^{LS}(3^1S_0\rightarrow 1^3S_1+1^3S_1)=\nonumber\\
&& -2\gamma
e^{-\frac{[m_1m_2(m_2-m_3)m_3+m^2_2m^2_3+m^2_1(m^2_2+m_2m_3+m^2_3)]P^2}{3\beta^2(m_1+m_3)^2(m_2+m_3)^2}}
 \sqrt{E_aE_bE_c}\frac{1}{\pi^{3/4}}(f_1+f_2) P\nonumber\\
&&\times \left
[45\beta^4(m_1+m_3)^4(m_2+m_3)^4(19m_1m_2+14m_1m_3+14m_2m_3+9m^2_3)
-12\beta^2(m_1+m_3)^2\right.\nonumber\\
&&~~~~\times
(m_2+m_3)^2(13m_1m_2+14m_1m_3+14m_2m_3+15m^2_3)(m_2m_3+2m_1m_2+m_1m_3)^2P^2\nonumber\\
&&~~~~\left. +4(m_2m_3+2m_1m_2+m_1m_3)^4(m_1m_2+2m_1m_3+2m_2m_3+3m^2_3)P^4 \right ]\nonumber\\
 &&\times\frac{1}{2187\sqrt{5}\beta^{11/2}}\frac{1}{(m_1+m_3)^5(m_2+m_3)^5}\\
 &&{\cal{M}}^{LS}(3^1S_0\rightarrow 1^3S_1+1^1S_0)=\nonumber\\
&& -\sqrt{\frac{2}{5}}\gamma
e^{-\frac{[m_1m_2(m_2-m_3)m_3+m^2_2m^2_3+m^2_1(m^2_2+m_2m_3+m^2_3)]P^2}{3\beta^2(m_1+m_3)^2(m_2+m_3)^2}}
 \sqrt{E_aE_bE_c}\frac{1}{\pi^{3/4}}(f_1-f_2) P\nonumber\\
&&\times\left[45\beta^4(m_1+m_3)^4(m_2+m_3)^4(19m_1m_2+14m_1m_3+14m_2m_3+9m^2_3)-12\beta^2(m_1+m_3)^2\right.\nonumber\\
&&~~~~\times(m_2+m_3)^2(13m_1m_2+14m_1m_3+14m_2m_3+15m^2_3)(m_2m_3+2m_1m_2+m_1m_3)^2P^2\nonumber\\
&&~~~~\left.+4(m_2m_3+2m_1m_2+m_1m_3)^4(m_1m_2+2m_1m_3+2m_2m_3+3m^2_3)P^4\right]\nonumber\\
&&\times \frac{1}{2187\beta^{11/2}}\frac{1}{(m_1+m_3)^5(m_2+m_3)^5}\\
&&{\cal{M}}^{LS}(3^1S_0\rightarrow 1^3P_0+1^1S_0)=2i\gamma
e^{\frac{-[(m_1m_2(m_2-m_3)m_3+m^2_2m^2_3+m^2_1(m^2_2+m_2m_3+m^2_3)]P^2}{3\beta^2(m_1+m_3)^2(m_2+m_3)^2}}\sqrt{E_aE_bE_c}\frac{1}{\pi^{3/4}}\nonumber\\
&&\times\left\{-(f_1+f_2)\frac{5\sqrt{5}}{81\sqrt{3}\beta^{1/2}}\right.\nonumber\\
&&~~+\left.\left [(16m^2_2m^2_3+14m_2m^3_3+17m^2_1m^2_2+17m^2_1m_2m_3+2m^2_1m^2_3+36m_1m^2_2m_3+37m_1m_2m^2_3+5m_1m^3_3)f_2\right.\right.\nonumber\\
&&~~~~+\left.\left.(2m^2_2m^2_3+5m_2m^3_3+17m_1m^2_2m_3+37m_1m_2m^2_3+14m_1m^3_3+17m^2_1m^2_2+36m^2_1m_2m_3+16m^2_1m^2_3)f_1\right]\right.\nonumber\\
&&~~~~\times\frac{\sqrt{5}}{243\sqrt{3}\beta^{5/2}}\frac{1}{(m_1+m_3)^2(m_2+m_3)^2}P^2\nonumber\\
&&~~+\left[(11m^2_2m^2_3+14m_2m^3_3+3m^2_1m^2_2+3m^2_1m_2m_3-3m^2_1m^2_3+16m_1m^2_2m_3+21m_1m_2m^2_3-m_1m^3_3)f_2\right.\nonumber\\
&&~~~~+\left.(3m^2_1m^2_2+16m^2_1m_2m_3+11m^2_1m^2_3+3m_1m^2_2m_3+21m_1m_2m^2_3+14m_1m^3_3-3m^2_2m^2_3-m_2m^3_3)f_1\right]\nonumber\\
&&~~~~\times\frac{-4}{729\sqrt{15}\beta^{9/2}}\frac{(m_2m_3+2m_1m_2+m_1m_3)^2}{(m_1+m_3)^4(m_2+m_3)^4}P^4\nonumber\\
&&~~+\left[(m_1m_2-m_1m_3+2m_2m_3)f_2+(m_1m_2+2m_1m_3-m_2m_3)f_1\right](m_2m_3+2m_1m_2+m_1m_3)^4\nonumber\\
&&~~~~\left.\times(m_1m_2+2m_1m_3+2m_2m_3+3m^2_3)\frac{4}{6561\sqrt{15}\beta^{13/2}}\frac{1}{(m_1+m_3)^6(m_2+m_3)^6}P^6\right\}\\
&&{\cal{M}}^{LS}(3^1S_0\rightarrow 1^3P_2+1^1S_0)=2i\gamma
e^{\frac{-[(m_1m_2(m_2-m_3)m_3+m^2_2m^2_3+m^2_1(m^2_2+m_2m_3+m^2_3)]P^2}{3\beta^2(m_1+m_3)^2(m_2+m_3)^2}}\sqrt{E_aE_bE_c}\frac{1}{\pi^{3/4}}\nonumber\\
&&\times\left\{\left [\left(-\frac{11m^2_1}{(m_1+m_3)^2}+\frac{14m_2(6m_2+5m_3)}{(m_2+m_3)^2}+\frac{m_1(28m_2+25m_3)}{(m_1+m_3)(m_2+m_3)}\right)f_2\right.\right.\nonumber\\
&&~~~~+\left.\left(\frac{14m^2_1}{(m_1+m_3)^2}+\frac{m_2(14m_2+25m_3)}{(m_2+m_3)^2}+\frac{m_1(73m_2+70m_3)}{(m_1+m_3)(m_2+m_3)}\right)f_1\right]\frac{-27\sqrt{2}}{6561\sqrt{15}\beta^{5/2}}P^2\nonumber\\
&&~~+\left [(12m^2_2m^2_3+14m_2m^3_3+7m^2_1m^2_2+7m^2_1m_2m_3-2m^2_1m^2_3+20m_1m^2_2m_3+23m_1m_2m^2_3-m_1m^3_3)f_2\right.\nonumber\\
&&~~~~+\left.(-2m^2_2m^2_3-m_2m^3_3+7m^2_1m^2_2+20m^2_1m_2m_3+12m^2_1m^2_3+7m_1m^2_2m_3+23m_1m_2m^2_3+14m_1m^3_3)f_1\right]\nonumber\\
&&~~~~\times(m_2m_3+2m_1m_2+m_1m_3)^2\frac{4\sqrt{2}}{729\sqrt{15}\beta^{9/2}}\frac{1}{(m_1+m_3)^4(m_2+m_3)^4}P^4\nonumber\\
&&~~+\left[(m_1m_2-m_1m_3+2m_2m_3)f_2+(m_1m_2+2m_1m_3-m_2m_3)f_1\right](m_2m_3+2m_1m_2+m_1m_3)^4\nonumber\\
&&~~~~\left.\times(m_1m_2+2m_1m_3+2m_2m_3+3m^2_3)\frac{-4\sqrt{2}}{6561\sqrt{15}\beta^{13/2}}\frac{1}{(m_1+m_3)^6(m_2+m_3)^6}P^6\right\}
\end{eqnarray}
}

\vspace*{-1cm}
\section*{Appendix B: Flavor and Weight factors}
\indent \vspace*{-0.5cm}

The flavor factors $f_1$ and $f_2$  can be calculated using the
matrix notation introduced in Ref.\cite{3p0rev3} with the meson
flavor wavefunctions following the conventions of
Ref.\cite{flavorfun} for the special process with definite charges
like $n\bar{n}\rightarrow \rho^+\rho^-$. In order to obtain the
general (i.e. charge independent) width of decays like
$n\bar{n}\rightarrow \rho\rho$, one should multiply the width
$\Gamma(n\bar{n}\rightarrow \rho^+\rho^-)$ by a charge
multiplicity factor ${\cal{F}}$. The $f_1$, $f_2$ and ${\cal{F}}$
for all the processes considered in this work are given in Table
2.
\begin{table}[hbt]
\begin{center}
\caption{\small Flavor and charge multiplicity factors}
\vspace*{0.5cm}
\begin{tabular}{ccccc}\hline\hline
General decay & subprocess& $f_1$& $f_2$ & ${\cal{F}}$\\\hline
$n\bar{n}\rightarrow\rho\rho$
&$n\bar{n}\rightarrow\rho^+\rho^-$
&$-\frac{1}{\sqrt{6}}$
&$-\frac{1}{\sqrt{6}}$
&$\frac{3}{2}$\\
$n\bar{n}\rightarrow a_0(1450)\pi$
&$n\bar{n}\rightarrow a_0(1450)^+\pi^-$
&$-\frac{1}{\sqrt{6}}$
&$-\frac{1}{\sqrt{6}}$
&$3$\\
$n\bar{n}\rightarrow a_2(1320)\pi$
&$n\bar{n}\rightarrow a_2(1320)^+\pi^-$
&$-\frac{1}{\sqrt{6}}$
&$-\frac{1}{\sqrt{6}}$
&$3$\\
$n\bar{n}\rightarrow\omega\omega$
&$n\bar{n}\rightarrow\omega\omega$
&$\frac{1}{\sqrt{6}}$
&$\frac{1}{\sqrt{6}}$
&$\frac{1}{2}$\\
$n\bar{n}\rightarrow K^\ast K$
&$n\bar{n}\rightarrow K^{\ast +}K^-$
&$-\frac{1}{\sqrt{6}}$
&$0$
&$4$\\
$n\bar{n}\rightarrow K^\ast K^\ast$
&$n\bar{n}\rightarrow K^{\ast +}K^{\ast -}$
&$-\frac{1}{\sqrt{6}}$
&$0$
&$2$\\
$s\bar{s}\rightarrow K^\ast K$ &$s\bar{s}\rightarrow K^{\ast
+}K^-$ &$0$ &$-\frac{1}{\sqrt{3}}$
&$4$\\
$s\bar{s}\rightarrow K^\ast K^\ast$

&$s\bar{s}\rightarrow K^{\ast+}K^{\ast -}$ &$0$
&$-\frac{1}{\sqrt{3}}$ &$2$\\\hline\hline
\end{tabular}
\end{center}
\end{table}


\baselineskip 18pt
\begin{thebibliography}{99}
\bibitem{x1835} M. Ablikim et al., BES collaboration, Phys. Rev. Lett. {\bf
95}, 262001 (2005)
\bibitem{pp} J. Z. Bai et al.,  BES Collaboration, Phys. Rev.
Lett. {\bf 91}, 022001 (2003)
\bibitem{g1} N. Kochelev, D. P. Min, Phys. Rev. D {\bf 72}, 097502 (2005); Phys. Lett. B {\bf 633},
283
(2006)
\bibitem{g2}X. G. He, X. Q. Li, X. Liu, J. P. Ma, Eur. Phys. J.
C {\bf 49}, 731 (2007)
\bibitem{g3} B. A. Li, Phys. Rev. D  {\bf 74}, 034019 (2006)
\bibitem{b1} G. J. Ding, M. L. Yan, Phys. Rev. C {\bf 72}, 015208
(2005)
\bibitem{b2} S. L. Zhu, C. S. Gao, Commun. Theor. Phys. {\bf
46},291 (2006); Z. G. Wang and S. L. Wan, J. Phys. G {\bf 34}, 505
(2007)
\bibitem{hz} T. Huang and S. L. Zhu, Phys. Rev. D {\bf 73},
014023 (2006)
\bibitem{KZ} E. Klempt and A. Zaitsev, Phys. Rept. {\bf 454}, 1
(2007)
\bibitem{MKo1} R. M. Baltrusaitis et al., MARKIII Collaboration,
Phys. Rev. Lett. {\bf 55}, 1723 (1985)
\bibitem{MKr1}R. M. Baltrusaitis et al., MARKIII Collaboration,
Phys. Rev. D {\bf 33}, 1222 (1986)
\bibitem{DM2r1} D. Bisello et al., DM2 Collaboration, Phys. Rev. D
{\bf 39}, 701 (1989)
\bibitem{DM2o1} D. Bisello et al., DM2 Collaboration, Phys. Lett.
B {\bf 192}, 239 (1987)
\bibitem{vij} J. Vijande, F. Fernandez, A. Valcarce, J. Phys. G
{\bf 31}, 481 (2005)
\bibitem{pageli} P. R. Page and X. Q. Li, Eur. Phys. J. C {\bf 1},
579 (1998)
\bibitem{Bugg} D. V. Bugg et al., Phys. Lett. B {\bf353}, 378
(1995)
\bibitem{BES17601}J. Z. Bai et al., BES Collaboration, Phys. Lett. B {\bf 472}, 207
(2000)
\bibitem{BES17602}M. Ablikim et al., BES Collaboration, Phys. Rev.
D {\bf 73}, 112007 (2006)
\bibitem{pdg2006} W.-M. Yao et al., J. Phys. G {\bf 33}, 1
(2006)
\bibitem{3p0rev1} A. Le Yaouanc, L. Oliver, O. Pene, J-C. Raynal,
Phys. Rev. D {\bf 8}, 2223 (1973); Phys. Rev. D {\bf 9}, 1415
(1974); Phys. Rev. D {\bf 11}, 1272 (1975); Phys. Lett. B {\bf
71}, 397 (1977); Phys. Lett. B {\bf 72},57 (1977).
\bibitem{3p0rev2} A. Le Yaouanc, L. Oliver, O. Pene, J-C. Raynal,
Hadron transitons in the quark model ( Gordon and Breach Science
Publishers, New York, 1988)
\bibitem{3p0rev3} W. Roberts and B. Silvestr-Brac, Few-Body Syst.
{\bf 11}, 171 (1992)
\bibitem{3p0rev4} H. G. Blundel, hep-ph/9608473
\bibitem{micu} L. Micu, Nucl. Phys. B {\bf 10}, 521 (1969)
\bibitem{3p00} S. Capstick, N. Isgur, Phys. Rev. D {\bf 34}, 2809
(1986); S. Capstick, W. Roberts, Phys. Rev. D {\bf 49} 4570 (1994)
\bibitem{3p0y} P. Geiger, E. S. Swanson, Phys. Rev. D {\bf 50},
6855 (1994)
\bibitem{3p0x}H.G. Blundell, S. Godfrey, Phys. Rev. D {\bf
53},3700 (1996)
\bibitem{3p0x1} H. G. Blundell, S. Godfrey, B.
Phelps, Phys. Rev. D {\bf 53}, 3712 (1996)
\bibitem{3p0x2} R. Kokoski, N. Isgur, Phys. Rev. D  {\bf 35}, 907
(1987)
\bibitem{3p01} E. S. Ackleh, T. Barnes and E. S. Swanson,
Phys. Rev. D {\bf 54}, 6811 (1996);
\bibitem{3p02} T. Barnes, F. E. Close, P. R. Page and E. S.
Swanson, Phys. Rev. D  {\bf 55}, 4157 (1997)
\bibitem{3p03} T. Barnes, N. Black and P. R. Page, Phys. Rev. D
{\bf 68}, 054014 (2003)
\bibitem{quarkmass}F. E. Close, E. S. Swanson, Phys. Rev. D
{\bf72}, 094004 (2005)
\bibitem{3p04} L. Burakovsky, P. R. Page, Phys. Rev. D {\bf 62}, 014011 (2000); H. Q. Zhou, R. G. Ping, B. S. Zou, Phys. Lett. B
{\bf 611}, 123 (2005); J. Lu, W. Z. Deng, X. L. Chen, S. L. Zhu,
Phys. Rev. D {\bf 73}, 054012 (2006); B. Zhang, X. Liu, W. Z.
Deng, S. L. Zhu, Eur. Phys. J. C {\bf 50}, 617 (2007); F. E.
Close, C. E. Thomas, O. Lakhina, E. S. Swanson, Phys. Lett. B {\bf
647}, 159 (2007); O. Lakhina, E. S. Swanson, Phys. Lett. B {\bf
650}, 159 (2007); C. Chen, X. L. Chen, X. Liu, W. Z. Deng, S. L.
Zhu, Phys. Rev. D {\bf 75}, 094017 (2007); G. J. Ding, M. L. Yan,
Phys. Lett. B {\bf657}, 49 (2007)

\bibitem{mock}C. Hayne and N. Isgur, Phys. Rev. D {\bf 25}, 1944
(1982)
\bibitem{recp} M. Jacob, G. C. Wick, Ann. Phys. {\bf 7}, 404
(1959)
\bibitem{jpglidm} D. M. Li, H. Yu and Q. X. Shen, J. Phys. G  {\bf
27}, 807 (2001)
\bibitem{epja-1} De-Min Li, Ke-Wei Wei and Hong Yu, Eur. Phys. J.
A {\bf 25}, 263 (2005)
\bibitem{pdg2002} K. Hagiwara et al., Phys. Rev. D {\bf 66},
010001 (2002)
\bibitem{okubo} S. Okubo, Prog. Theor. Phys. {\bf 27}, 949 (1962)

\bibitem{flavorfun} S. Godfrey, N. Isgur, Phys. Rev. D  {\bf 32},
189 (1985)
\end{thebibliography}
\end{document}